\documentclass[pra,twocolumn,showpacs,floatfix,amsmath,amssymb]{revtex4}
\usepackage{graphicx}
\usepackage{dcolumn}
\usepackage{bm}
\usepackage{float}
\begin{document}
\title{Bose-Einstein condensates with attractive 1/r interaction: The case of self-trapping}
\author{I. Papadopoulos, P. Wagner, G. Wunner,}
\email{wunner@theo1.physik.uni-stuttgart.de}
\author{J. Main}
\affiliation{%
1. Institut f\"ur Theoretische Physik,
         Universit\"at Stuttgart, D-70550 Stuttgart, Germany}
\date{\today}
\begin{abstract}
Amplifying on a proposal by O'Dell  et al. for the  realization of Bose-Einstein condensates of neutral atoms 
with attractive $1/r$ interaction, we point out that the instance of self-trapping of the condensate, without 
external trap potential, is physically best understood by introducing  appropriate "atomic" units. This reveals
a remarkable scaling property: the physics of the condensate depends only on the two  parameters $N^2 a/a_u$ and $\gamma/N^2$, where $N$ is the 
particle number, $a$ the scattering length, $a_u$ the "Bohr" radius and $\gamma$ the trap frequency in atomic units. We calculate accurate numerical results
for  self-trapping wave functions and potentials, for energies, sizes and peak densities, and compare
with previous variational results. As  a novel feature we point out the existence of a second solution of the extended 
Gross-Pitaevskii equation for negative scattering lengths, with and without 
trapping potential, which is born together with the ground state in a tangent bifurcation.  This indicates the existence of an 
unstable collectively excited state of the condensate for negative scattering lengths. 
\end{abstract}
\maketitle
\section{Introduction}
Bose-Einstein condensation of {\em dipolar} gases has attracted much attention in recent years \cite{santos00,baranov02,goral02a,goral02b,giovanazzi03} because it offers the opportunity to create
degenerate
quantum gases with adjustable long- and short-range interactions,
which gives rise to a wealth of novel phenomena \cite{giovanazzi02,santos03,li04,dell04}. In particular,
the achievement of Bose-Einstein condensation in a gas of chromium atoms
\cite{griesmaier05}, with a large dipole moment, has opened the way to promising experiments on 
dipolar quantum gases \cite{stuhler05}. 

As an alternative system with tunable interactions, the Bose-Einstein condensation of neutral atoms with electromagnetically induced attractive
$1/r$ interaction has been proposed. Here a  {\em monopolar}, "gravity-like", long-range interaction, in addition to the short-range (van-der-Waals-like) interactions, takes the place of the dipole-dipole interaction in dipolar
gases. A monopolar quantum gas could be realized according to 
O'Dell et al. \cite{dell00} by a combination of 6 appropriately 
arranged "triads" of intense off-resonant laser beams. In that arrangement, the  $1/r^3$ interactions 
of  the retarded dipole-dipole  interaction of neutral atoms in the presence of intense electromagnetic radiation are averaged out in the near-zone limit 
\cite{thirunamachandran80,craig84}, while  the weaker $1/r$ interaction is retained. The resulting
atom--atom interaction potential in the near-zone is \cite{dell00}
\begin{equation} {
V_u(\vec r, \vec r^\prime \,) = - \frac{u}{|\vec r - \vec r^\prime |}, \quad
\hbox{\rm  with} \quad u =
 \frac{11}{4\pi}\,  \frac{I k^2  \alpha^2}{c \varepsilon_0^2} 
} \, .\nonumber
\end{equation}
Here, $\alpha(k)$ is the isotropic, dynamic, polarizability of the 
atoms at frequency $c k$, and $I$ the intensity of the radiation. The
quantity $u$ determines  the strength of the "gravity-like" interaction.
The estimate for $u$ given by O'Dell et al. \cite{dell00} for CO$_2$ laser light of  intensity $I = 10^8$~W/cm$^2$ is equivalent to the attraction of two
opposite equal charges with $q  \approx e/2000$. However, by contrast with
the van der Waals interaction, the $1/r$ potential 
acts over the entire sample, and therefore its contribution to the energy can become important. Instead of 6 triads of lasers a different arrangement with three  rotating lasers has been proposed \cite{schuette07}.

Even though the experimental realization of such configurations is not
yet at hand, the theoretical issues  associated with monopolar degenerate quantum gases are
worthwhile investigating. In particular, as pointed out by O'Dell et al. \cite{dell00}, the intriguing new physical feature that emerges  is the possibility of {\em self-trapping} of the condensate, 
without external trap.  

In the theory of trapped Bose-Einstein condensates it is common
to introduce
as natural units for energy and length the quantum energy $\hbar \omega_0$ 
and the oscillator length $a_0= \sqrt{\hbar/m \omega_0}$ of the trap potential.
In the case of self-trapping, however, where the trapping potential is switched off,
$\hbar \omega_0 \to 0$ and  $a_0 \to \infty$.  Thus these 
quantities become  "bad" units. As a consequence, in their study of
the physical conditions necessary to observe the transition from external binding to self-binding Giovannazzi et al. \cite{giova01} used the laser wavelength and energy as units of length and energy. 

It is the purpose of this paper to re-analyze Bose condensates with attractive
$1/r$ interaction
using appropriate
"atomic" units. This will first reveal remarkable scaling properties of the
condensates. Next we solve the extended Gross-Pitaevskii equation for
monopolar quantum gases numerically
and compare with previous variational results. Last, as a novel feature, we point out that our numerical calculations reveal the existence of a second solution
of the Gross-Pitaevskii equation for negative scattering lengths, which is born together with the ground state in a bifurcation "out of nowhere". The existence of the second solution indicates the existence of an unstable
collectively excited state of such condensates at negative scattering lengths.

\section{Natural units, scaling properties}
\subsection{The general case}
We  analyze the physics of trapped monopolar gases, and in particular the limit $\omega_0 \to 0$, 
in terms of  natural "atomic" units. From the analogy 
$u \Leftrightarrow {e^2}/{4\pi \varepsilon_0}$ we can define a "fine-structure
constant" 
\begin {equation}
\alpha_u := {u}/{\hbar c}
 \; ,
\end{equation}
and  can construct  a "Bohr radius" and "Rydberg energy"
in the usual way from the Compton wavelength ${\mathchar '26\mkern -9mu\lambda}_{\rm C} = \hbar/mc$  and the rest energy $mc^2$ via
\begin{equation}
a_u = \frac{{\mathchar '26\mkern -9mu\lambda}_{\rm C}}{\alpha_u 
} = \frac{\hbar^2}{m u} \; , \quad 
E_u = \frac{\alpha_u^2 m c^2}{2} = \frac{\hbar^2}{2m a_u^2}\; .  
\end{equation}        
Measuring lengths in $a_u$ and energies in $E_u$, we can write the
 Hartree equation of the ground state of a system of $N$ identical bosons in an isotropic external trapping potential $V_0(r) = m \omega_0^2 r^2/2$, all in the same single-particle orbital $\psi$, 
interacting via $V_u$ and the $s$-wave scattering pseudopotential $V_s = 4 \pi a \hbar^2  \,\delta(\vec r - {\vec r\,}^\prime)/m$  in dimensionless form
\begin{eqnarray}\label{HFat}
\Big[ \hskip -1mm - \Delta + \gamma^2 r^2 + N 8 \pi \frac{a}{a_u} |\psi(\vec r)|^2 
\hskip -2mm
&-& \hskip -2mm 2 N  \hskip -1mm  \int \frac{|\psi({\vec r\,}^\prime)|^2}{|{\vec r}- {\vec r\,}^\prime |}
d^3 {\vec r\,}^\prime \Big] \; \psi(\vec r)  \nonumber \\[1.1ex]
 &=& \varepsilon \, \psi(\vec r) \, . 
\end{eqnarray}

In (\ref{HFat}), $\varepsilon$ is the chemical potential, and the dimensionless quantity $\gamma$  denotes the quantum energy of the trapping frequency in units of the  "Rydberg" energy
\begin{equation}
\gamma = \hbar \omega_0/E_u.  
\end{equation} 
Small values of $\gamma$ imply that the effects of the trapping potential are small 
compared with the effects of the gravity-like interaction, and vice versa for 
large values of $\gamma$. In (\ref{HFat})
we have also assumed $N\gg1$ so that the usual prefactor $(N-1)$ in the Hartree
potential can be replaced with the total particle number $N$. Using the
"order parameter"  $ \Psi = \sqrt{N}\psi$ instead of the single-particle orbital, one can absorb
the $N$-dependence in (\ref{HFat}) in the wave function $\Psi$, and 
obtains an extended 
(or, for vanishing gravity-like interaction, the familiar) time-independent
Gross-Pitaevskii equation. 

From (\ref{HFat}) it would seem that there are three physical parameters
governing the problem: the trap frequency $\omega_0$, given by the dimensionless
quantity $\gamma$, the particle number $N$ and the relative strength $a/a_u$
of the scattering  and the gravity-like potential. 
For the example mentioned before 
one has an estimate of $a \sim 10^{-9}$~m, $a_u \sim 2.5 \cdot 10^{-4}$, thus
$a/a_u \sim 10^{-6} - 10^{-5}$. 

However, a central result of the present 
paper is that the physics of degenerate monopolar gases  depends only on 
 {\em two} relevant parameter, viz. $\gamma/N^2$ and $N^2 a/a_u$. To see this we note a
remarkable scaling property of the mean-field Hamiltonian in  (\ref{HFat}): Let $\psi(\vec r)$ be a solution of the (formal) {\em one}-boson problem for a given scaling length $a/a_u$ and trap frequency $\gamma$,
\begin{equation}
H_{{\rm mf}}{(N=1, a/a_u, \gamma)}(\vec r) \; \psi(\vec r) = \varepsilon \; \psi(\vec r)
\end{equation}
then {$\tilde \psi\, := \,N^{3/2} \, \psi(\vec{\tilde r})$}, with 
{$\vec{\tilde r} = {\vec r}/N$}, solves the {$N$}-boson problem for the  
scaled scattering length {$N^2\, a/a_u$} and the scaled trap frequency $ \gamma/N^2$:
\begin{eqnarray}\label{scaling}
H_{{\rm mf}}{(N, N^2a/a_u, \gamma/N^2)} \,(\tilde{\vec r}\,) \; \tilde \psi\,(\tilde{\vec r}\,)  &=& {\tilde \varepsilon} \; {\tilde \psi}\,(\tilde{\vec r}\,) \nonumber \\[1.1ex]
\quad \hbox{\rm with } 
{\tilde \varepsilon} &= & N^2 \varepsilon \, .
\end{eqnarray}
The proof is straightforward and left to the reader.
From (\ref{scaling}) follow scaling properties for the mean-field energy,
the root-mean square radius of the condensate and its peak density, respectively:
\begin{eqnarray}
 E (N, N^2 a/a_u, \gamma/N^2) &=& N^3 \,E(N=1, a/a_u, \gamma) \nonumber \\
\sqrt{\langle r^2 \rangle}_{|{(N, N^2a/a_u, \gamma/N^2)}} &=& \, \sqrt{\langle r^2 \rangle}_{|(N=1, a/a_u,\gamma)} / {\sqrt{N}}  \nonumber \\
 \varrho_{|{(N, N^2a/a_u), \gamma/N^2})} &=& N^4 \varrho_{|(N=1, a/a_u, \gamma/N^2)} \nonumber \\
&=& N^4 \, |\psi(0)|^2{.~~~~}
\end{eqnarray}

Iso-surfaces with constant $N^2 a/a_u$ and $\gamma/N^2$ form planes in the
three-dimensional parameter space 
$(\gamma, N, a/a_u)$ on a logarithmic scale.

\subsection{Self-binding}

\begin{figure}
\includegraphics[width=0.40\textwidth]{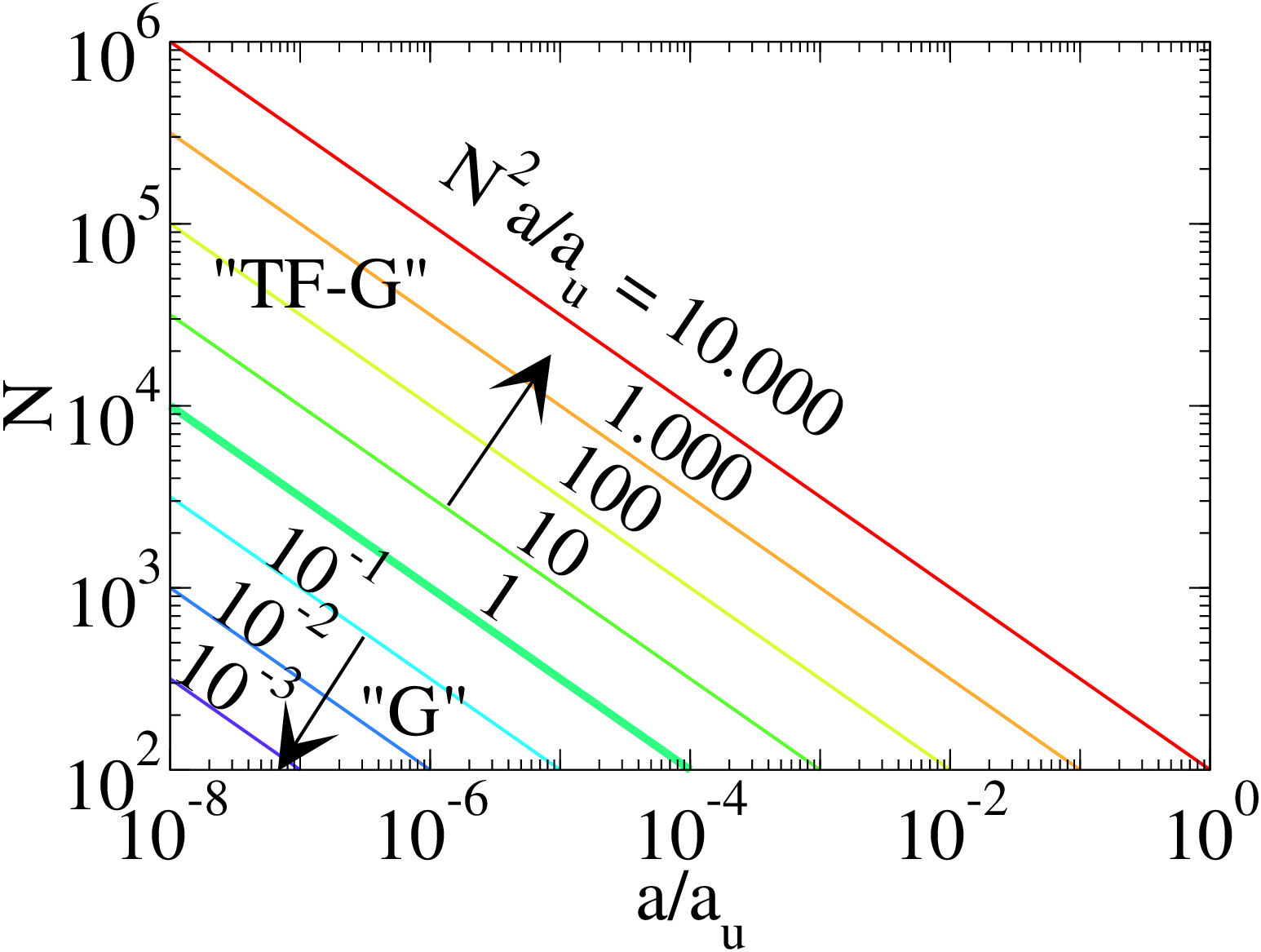}
\caption{\label{fig1} (Color online) Phase diagram $N$ versus $a/a_u$ for the self-binding ground state of monopolar degenerate quantum gases. (Explanation see text.)}
\end{figure}
In the case of self-binding we are left with one relevant parameter, $N^2a/a_u$.
In Fig.~\ref{fig1} we show the phase diagram $N$ vs. $a/a_u$ for self-binding degenerate monopolar quantum gases. 
Note that because of the scaling property, physics is  
identical on each of the sloping straight lines with $N^2a/a_u$ = constant.
Apart from a numerical factor, the relevant parameter $N^2a/a_u$ is identical to
the quantity $\tilde{s} \tilde{u}$ used by O'Dell et al. \cite{dell00}, but the universal nature 
of this quantity was not emphasized.
The two asymptotic regimes of self-trapping dubbed 
  "G" ("gravity") and "TF-G" ("Thomas-Fermi gravity") in \cite{dell00} are characterized 
by the size of the scaling parameter. For $N^2 a/a_u \gg 1$, the
kinetic energy is negligible, and self-binding results from the balance 
between repulsive scattering and  gravity-like attraction. For $N^2 a/a_u 
\ll 1$ scattering is negligible, and self-trapping appears by a balance
between kinetic energy and gravity-like attraction.

We note that the "G" regime 
corresponds to the Newton-Schr{\"od}inger scheme of quantum mechanics, which 
is  a nonlinear variant of  quantum mechanics that has been investigated in  detail in numerous publications on quantum measurement 
\cite{diosi84,bonilla91,bonilla92,jones95,kumar00,soni02,geszti04,greiner06}. 
In that regime, the extended time-independent Gross-Pitaevskii equation (\ref{HFat}) turns into  the Schr\"odinger equation of gravitationally self-interacting quantum particles. It is worth
noting that interacting monopolar quantum gases offer an experimental
realization of Newton-Schr\"odinger quantum mechanics.

\section{Numerical solution, results and discussion}

We have determined numerically accurate 
radially symmetric solutions of the extended Gross-Pitaevskii equation (\ref{HFat}) in dependence on the 
scaling parameter $N^2 a/a_u$ both for the self-binding case $\gamma/N^2 = 0$ and for $\gamma/N^2 \ne 0$.
To verify the numerical results two different methods were employed. One was to integrate in parallel 
equation (\ref{HFat}) and the Poisson equation for the gravity-like interaction numerically outward
from $r = 0$  by exploiting the initial conditions for
the first derivatives and setting initial values at $r = 0$ for the wave function $\psi_0$ and the effective potential
produced by the gravity-like interaction $V_0$. The latter was varied via bisection until convergence of 
the wave function to zero at large values of $r$ was attained. 
The other method was an iterative one: the wave functions determined
in the preceding step are used to calculate the effective potential in the next step, and the resulting one-dimensional 
Schr\"odinger equation is integrated until self-consistency is achieved.  The iteration is initialized by a
reasonable guess for the wave function.

\begin{figure}
\includegraphics[width=0.40\textwidth]{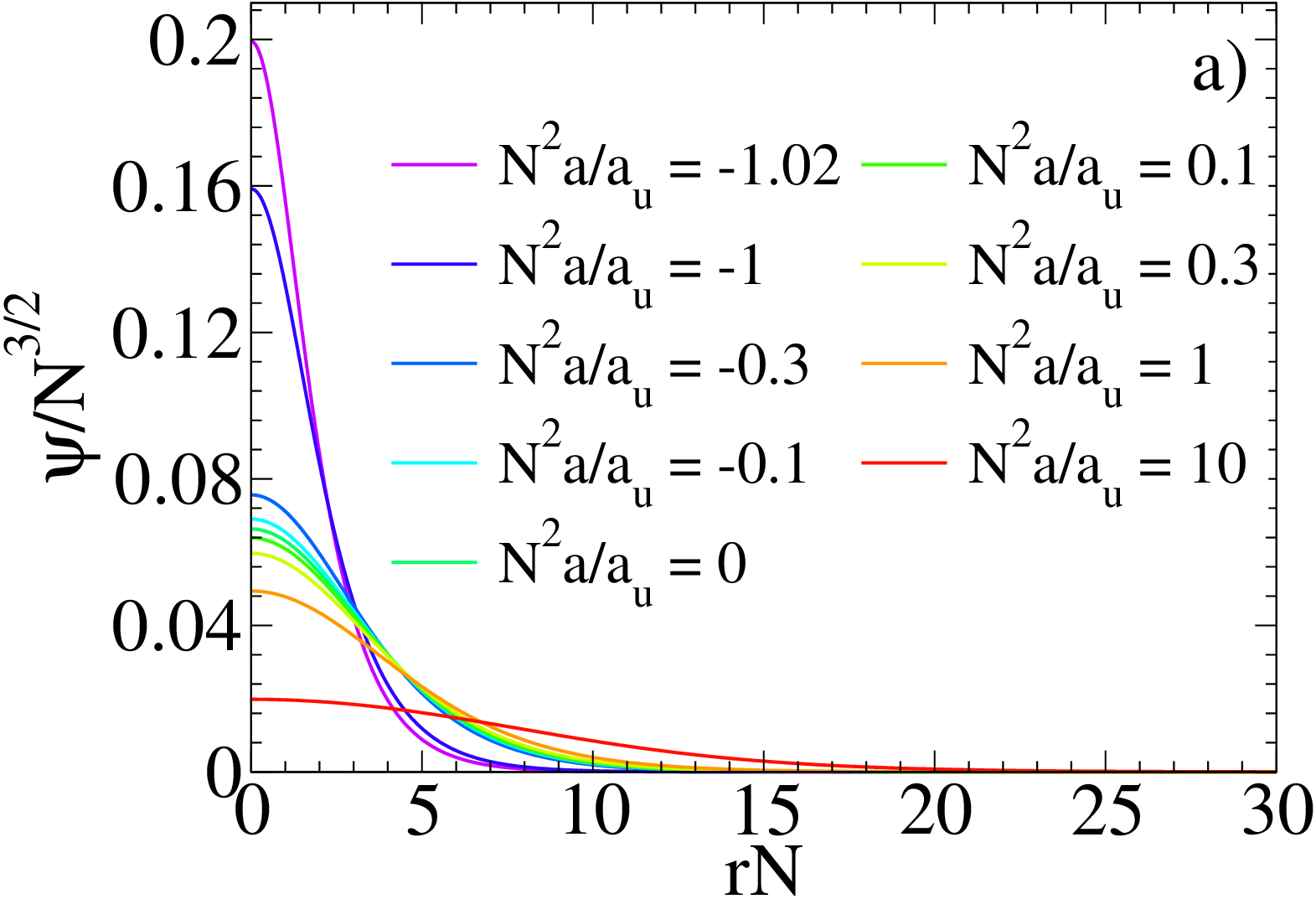}
\\[4ex]
\includegraphics[width=0.40\textwidth]{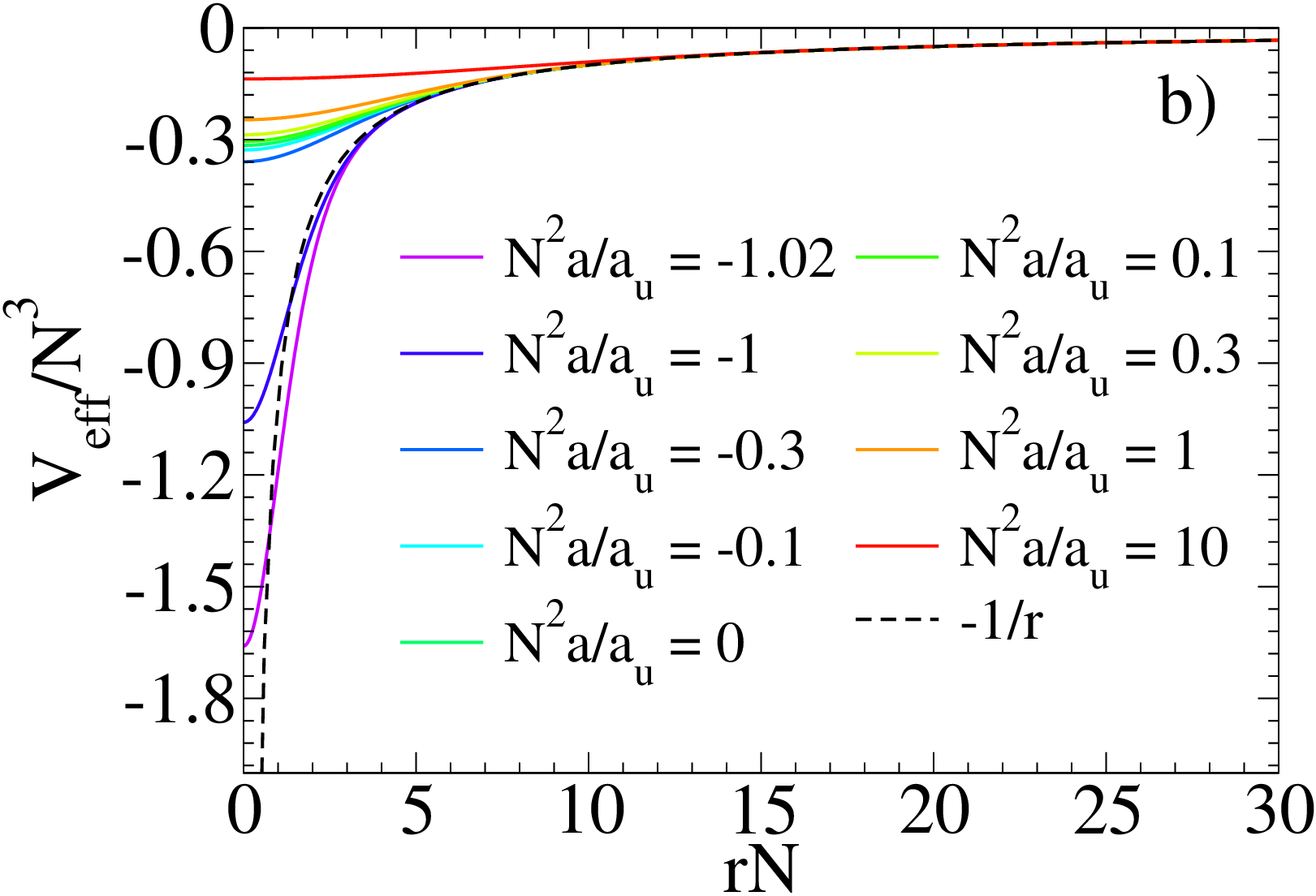}
\caption{\label{fig2} (Color online) Numerically accurate self-binding ground state s-wave solutions for different values of the scaling parameter $N^2 a/a_u$\,: a) wave functions;
b)  self-binding potentials. Both in a) the value of the wave function at the origin 
and in b) the absolute value of the self-binding potential at the origin decrease monotonically with  the scaling parameter, from their
maximum values at $N^2 a/a_u = -1.02$ to their smallest values at $N^2 a/a_u =  10$. Thus, as  the scaling parameter grows the binding becomes weaker. In b) the asymptotic $1/r$ potential is also shown for comparison.
}
\end{figure}

In Fig.~\ref{fig2} we 
show our results for the 
wave functions and the corresponding self-consistent potentials for the case of self-binding for 
different values of the scaling parameter $N^2 a/a_u$. It can be seen that for increasing
$N^2a/a_u$ the potentials grow shallower and the wave functions become more
extended. The figure confirms that asymptotically all self-binding 
potentials converge to a $1/r$ potential \cite{greiner06}. The case of $N^2 a/a_u = 0$ corresponds to the solutions of the Newton-Schr{\"od}inger
equation \cite{greiner06,moroz98,moroz99,bernst98}. As already pointed out by O'Dell et al. \cite{dell00}
solutions also exist for negative scattering lengths, where the contact interaction, in addition to the gravity-like 
interaction, becomes attractive and stability of the condensate is established by the equilibrium of 
the kinetic energy of the condensate and the two attractive interactions. Fig.~\ref{fig2} shows that
for negative scattering lengths
the self-trapping potentials become ever more binding, until at a value of $N^2 a/a_u \approx
-1.0251$ no solutions can be found any more, and the condensate becomes unstable
with respect to collapse. This corrects the variational value  
of $N^2 a/a_u = -3 \pi/8 \approx -1.18$ given by O'Dell et al. \cite{dell00}. In their variational
calculation, a Gaussian type orbital was assumed, and the mean-field energy of the condensate
was minimized with respect to the width of the Gaussian. 

\begin{figure}
\includegraphics[width=0.35\textwidth]{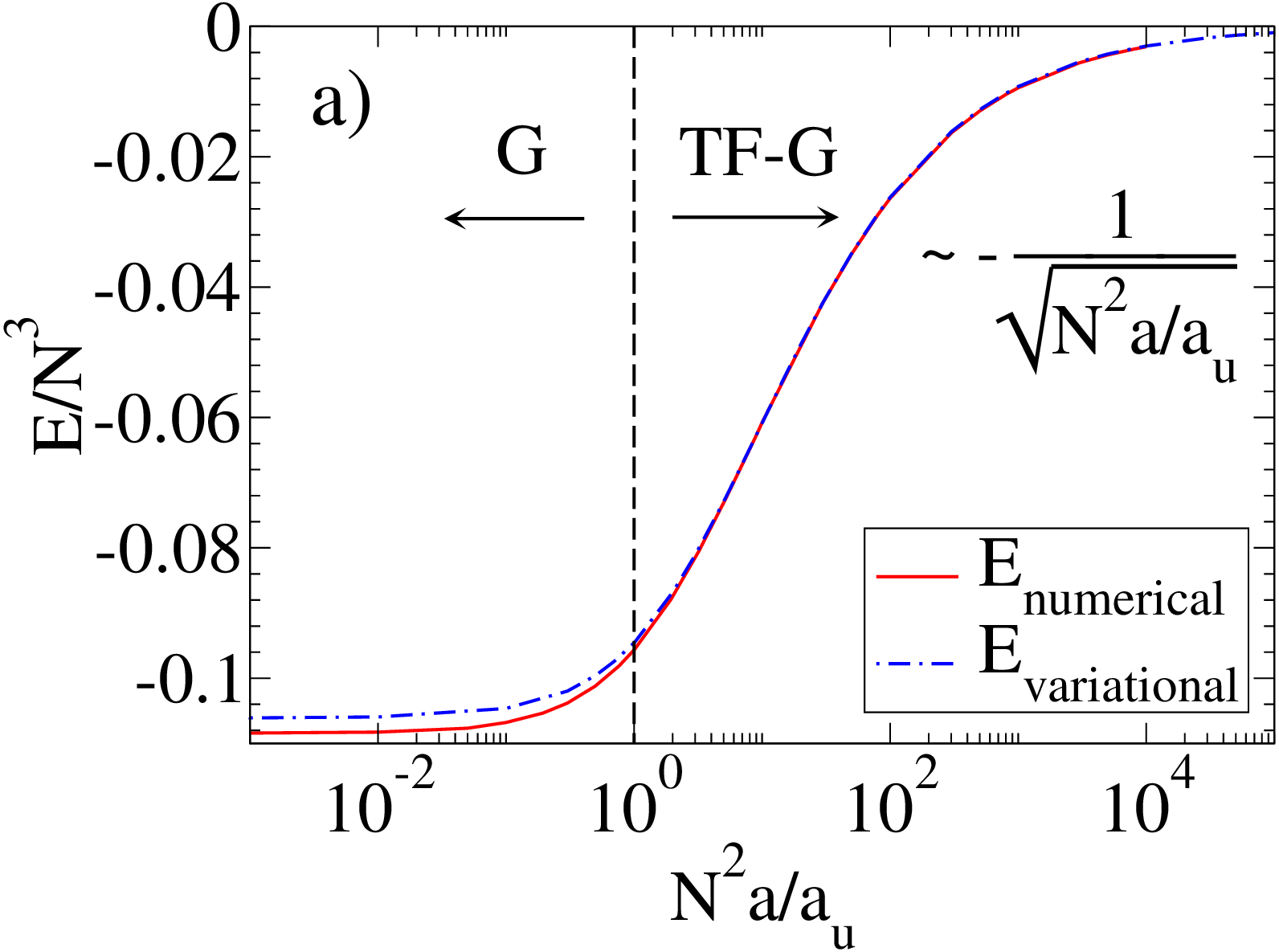}
\\[1.7ex]
\includegraphics[width=0.35\textwidth]{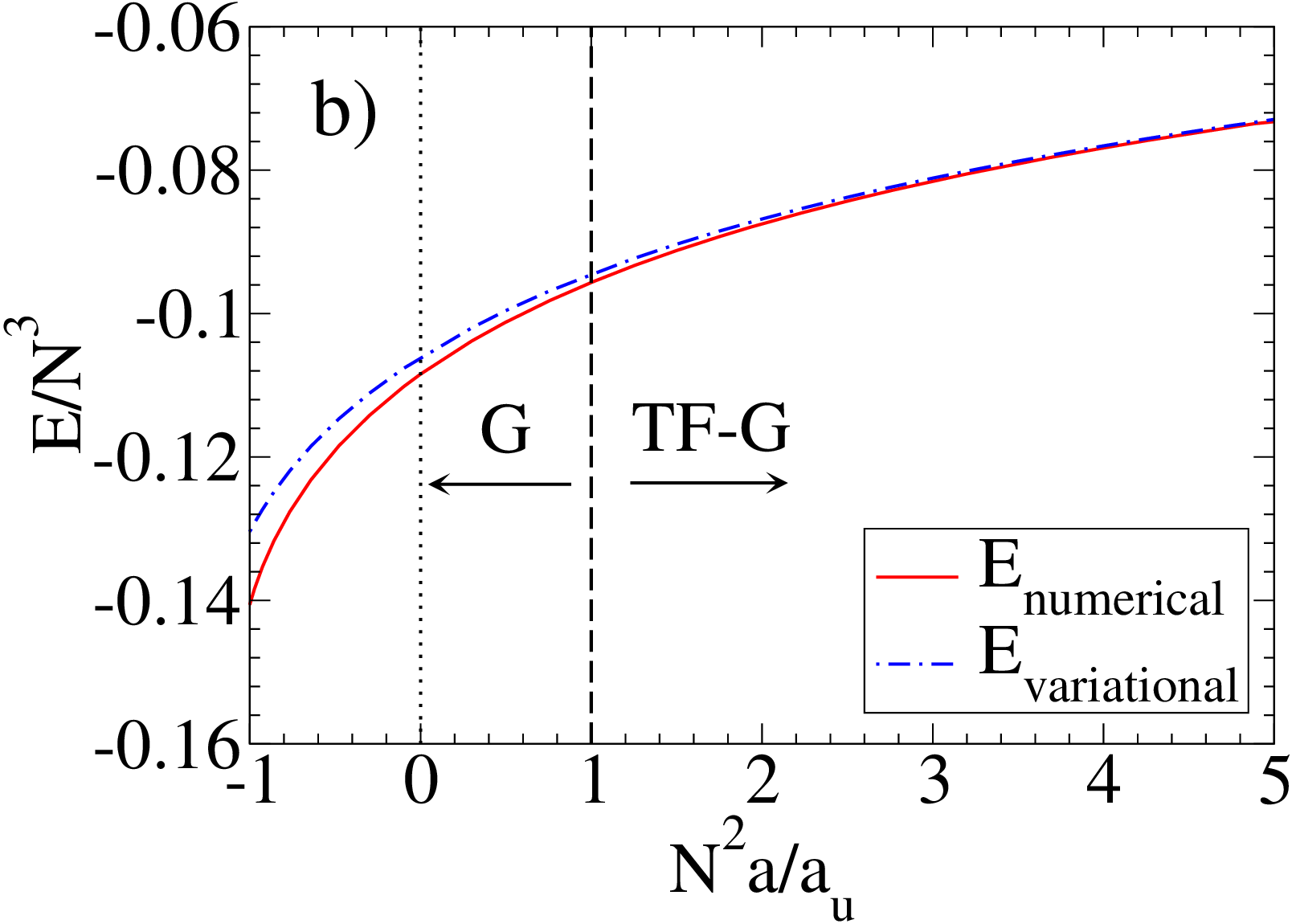}
\caption{\label{fig3} (Color online)  Total energy of the condensate as a function of $N^2 a/a_u$\,: a) on a logarithmic  and b) on a linear scale. Variational results obtained by minimizing the total energy for a Gaussian type orbital \cite{dell00} are 
shown by dashed lines.
}
\end{figure}

Since we have the numerically accurate solutions at hand, we are in a position to check 
the accuracy of the variational results for observables of the condensate  obtained 
by O'Dell et al. \cite{dell00}.  
Fig.~\ref{fig3} shows the behavior of the total energy of the condensate over seven decades of 
the scaling parameter $N^2 a/a_u$. To cover the range of negative scattering lengths, the energy is
also given  on a linear scale in the range around $N^2 a/a_u \approx 0$. The transition between 
the two asymptotic
regimes G and TF-G  around $N^2 a/a_u  \sim 1$ is evident
from Fig.~\ref{fig3}. The comparison with the variational results also plotted in Fig.~\ref{fig3} shows that the
TF-G regime is well described by the variational calculation. It is only in the transition to the
G regime, and in particular for negative values of $N^2 a/a_u$, that sizeable deviations can be observed,
up the order of 10 per cent. 

\begin{figure}
\includegraphics[width=0.35\textwidth]{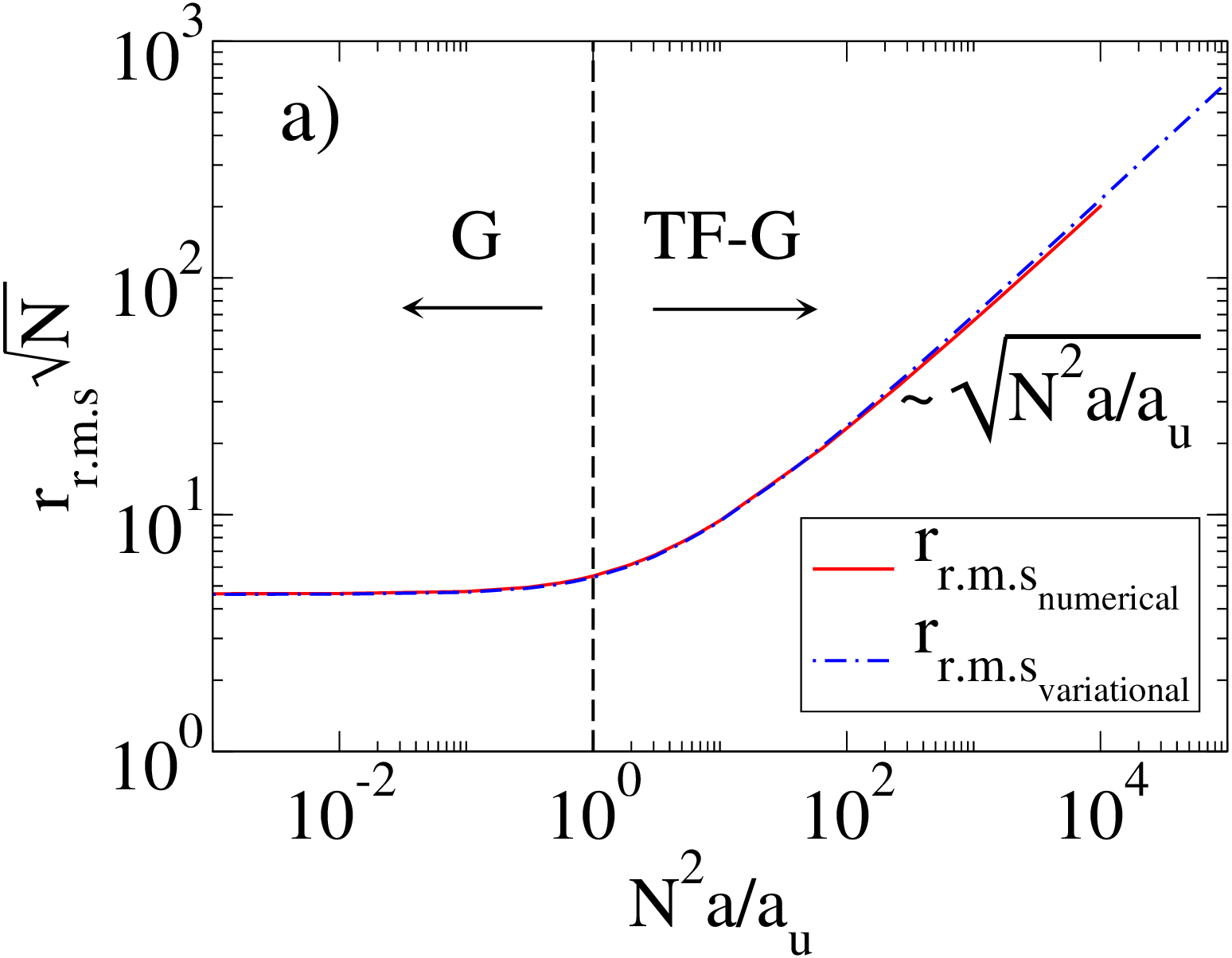}
\\[1.7ex]
\includegraphics[width=0.35\textwidth]{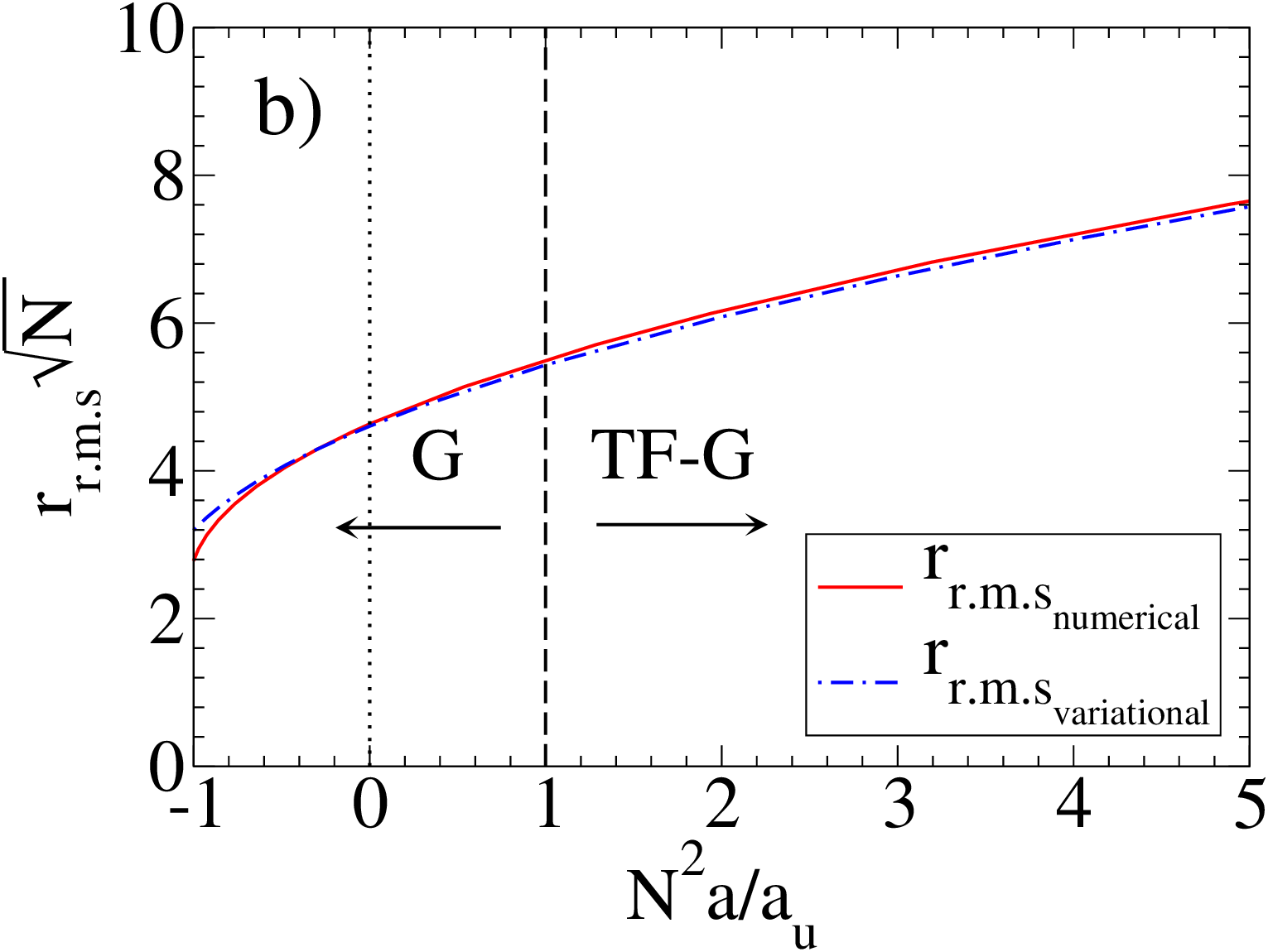}
\caption{\label{fig4} (Color online)  Root-mean-square radius  of the condensate as a function of $N^2 a/a_u$\,: a)  on a logarithmic  and b) on a linear scale. Variational results \cite{dell00} are 
shown by dashed lines.
}
\end{figure}

\begin{figure}
\includegraphics[width=0.35\textwidth]{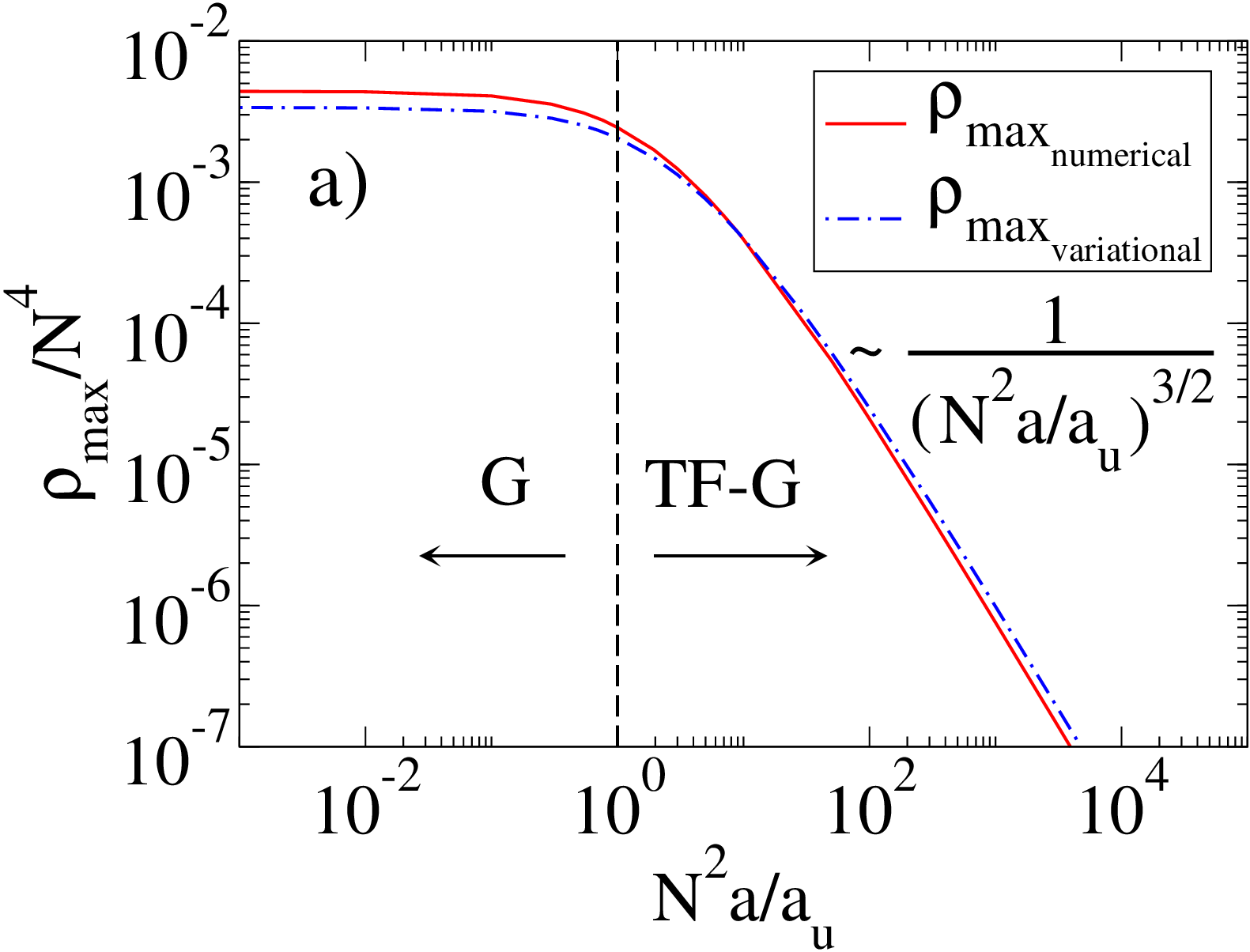}
\\[1.7ex]
\includegraphics[width=0.35\textwidth]{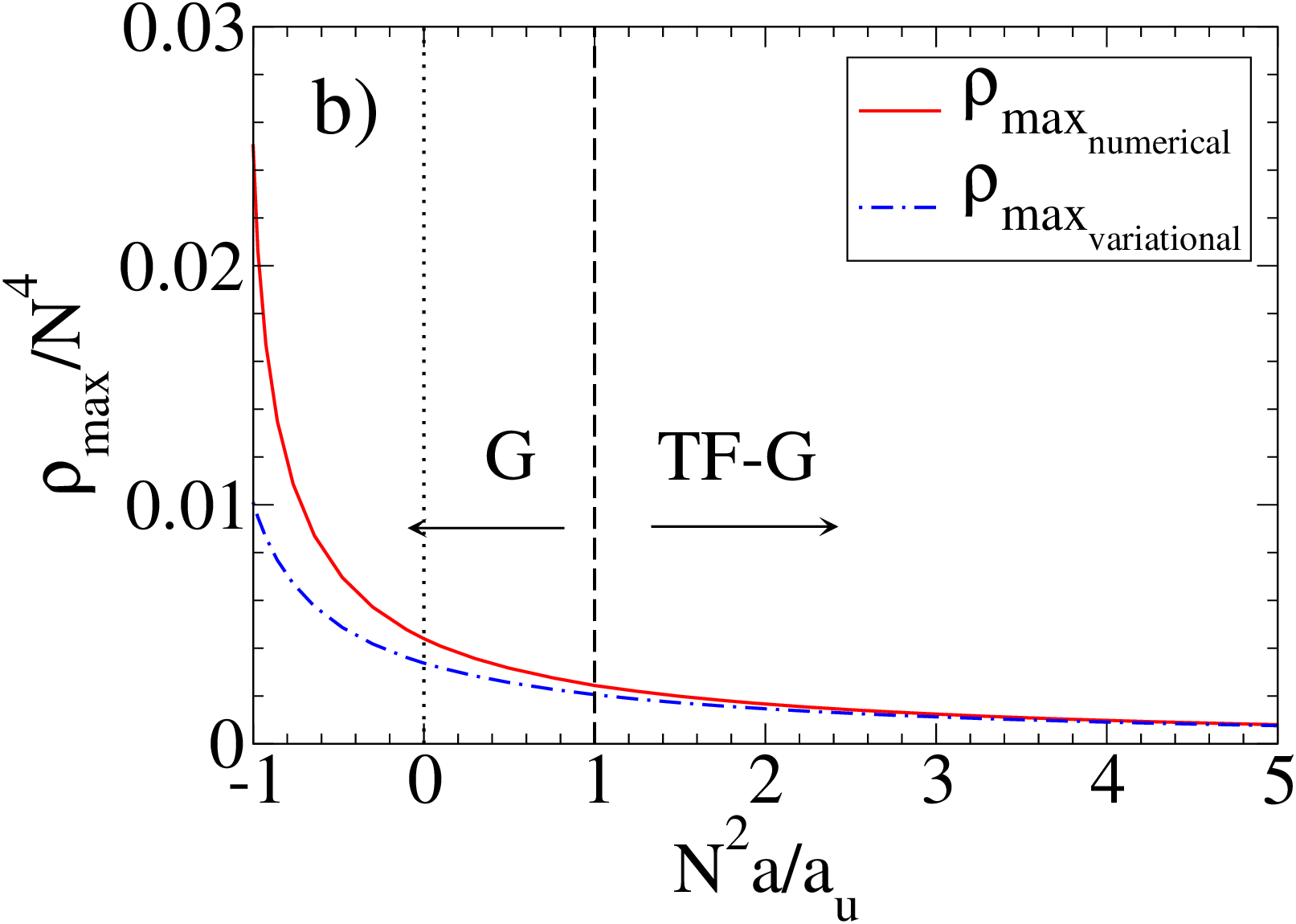}
\caption{\label{fig5} (Color online)  Peak density $\varrho$  of the condensate as a function of $N^2 a/a_u$\,: a) on a logarithmic  and b) on a linear scale. Variational results \cite{dell00} are 
shown by dashed lines.
}
\end{figure}

Observables other than the energy are  more sensitive to the accuracy of the wave function. 
We therefore compare our numerically accurate results with the variational results for the root-mean-square radius and 
the peak density of the condensate in Fig.~\ref{fig4} and Fig.~\ref{fig5}, again over seven decades of 
the scaling parameter on a logarithmic scale, and on a linear scale around  $N^2 a/a_u \approx 0$. Again the
transition between the two asymptotic regimes can be seen. It can also be recognized that the variational 
results well reproduce the overall behavior of the observables. For the extension of the condensate sizeable 
deviations occur again in the transition to the G regime and for  negative values of $N^2 a/a_u$, while for
the peak density the variational calculation overestimates the correct values in the TF-G regime, and underestimates
them in the G regime, and for negative scattering lengths. Here the deviations increase up to more than  
100 per cent. This is understandable since the peak density depends crucially on the correct wave function. 

\begin{figure}
\includegraphics[width=0.36\textwidth]{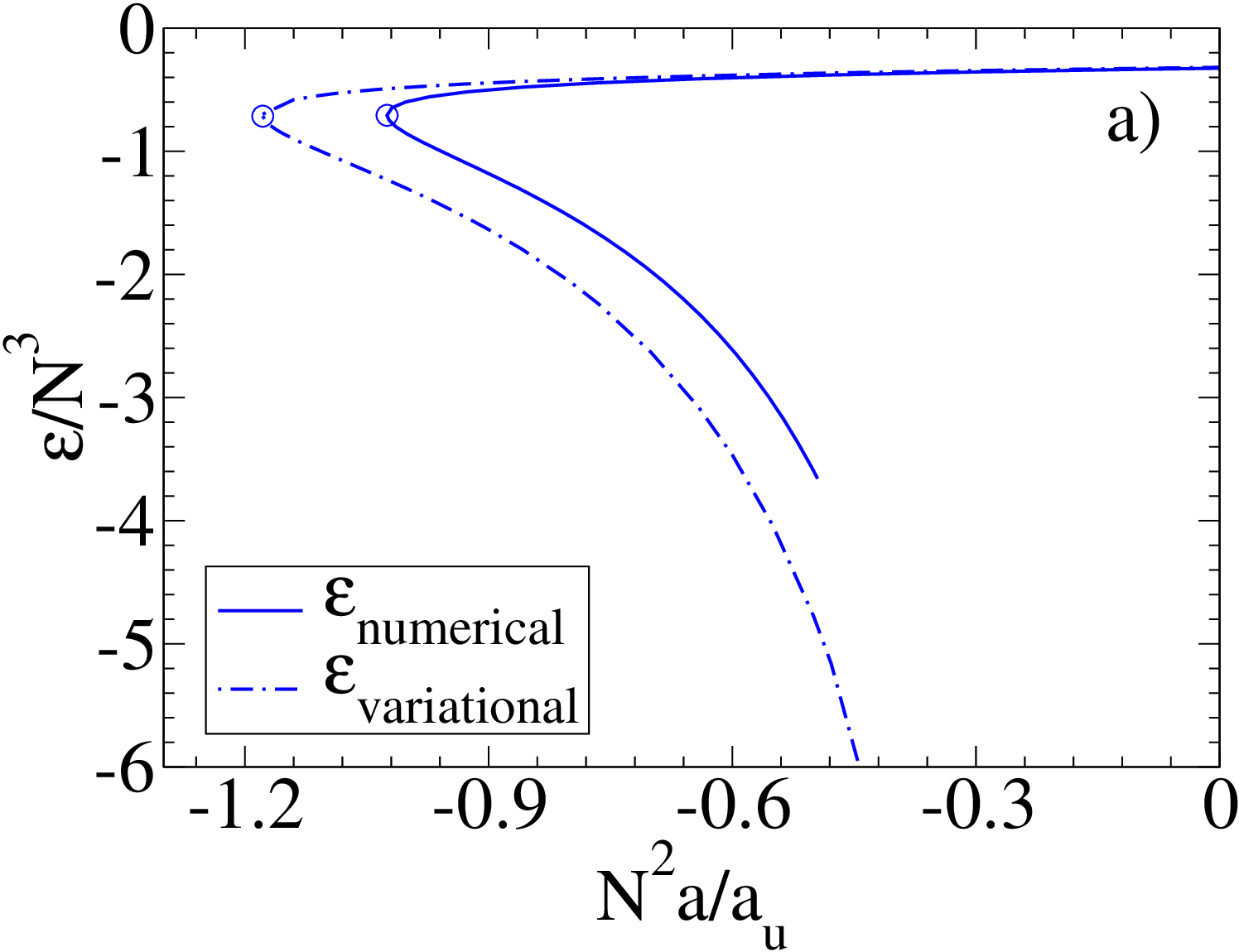}
\\[3.3ex]
\includegraphics[width=0.36\textwidth]{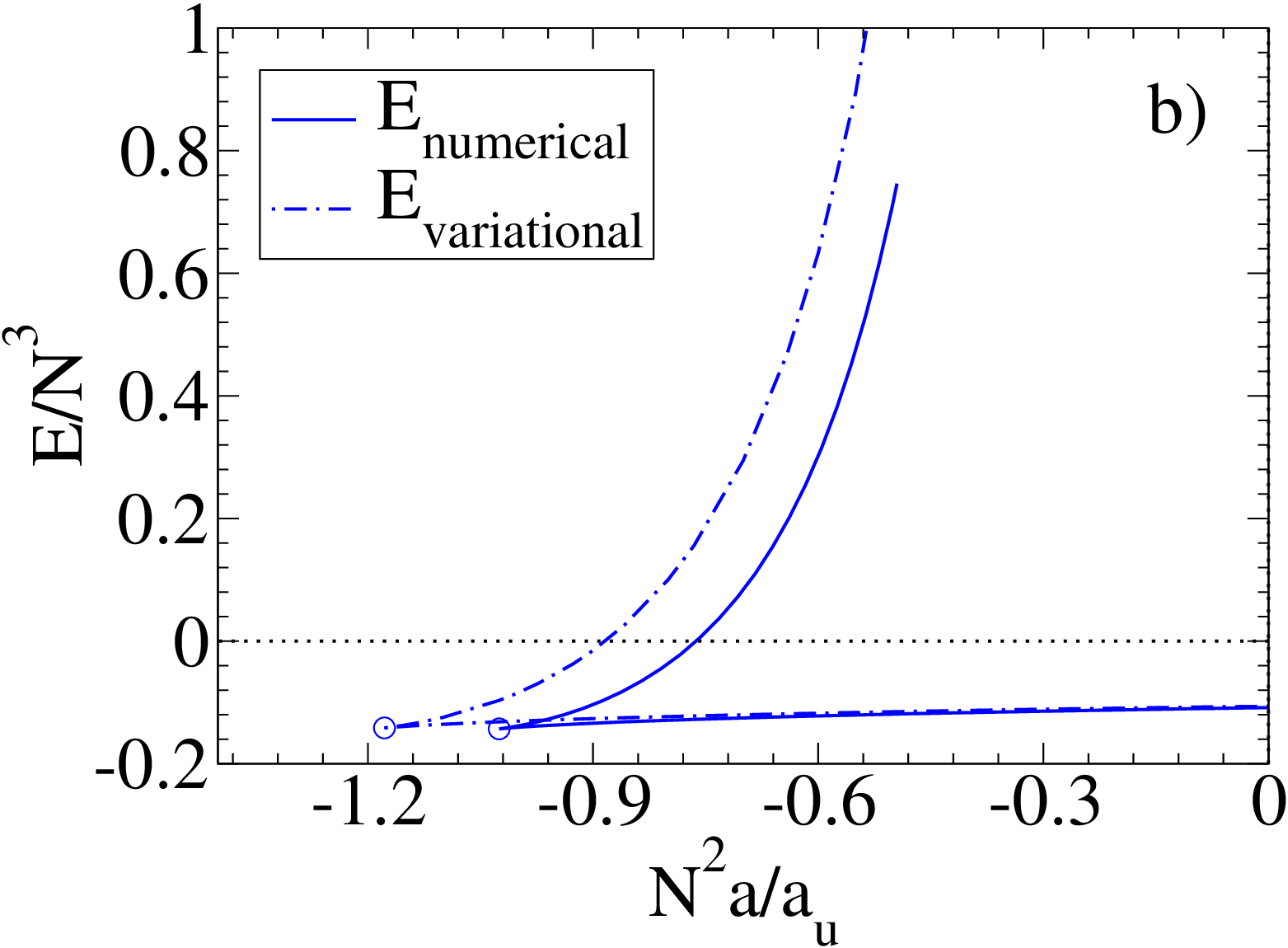}
\caption{\label{fig6} (Color online) a) Bifurcation of the chemical potential  and b) bifurcation of the 
total mean-field energy at the critical point $N^2 a/a_u = -1.0251$, for self-binding, i.~e., vanishing trap potential.
}
\end{figure}

\begin{figure}
\includegraphics[width=0.39\textwidth]{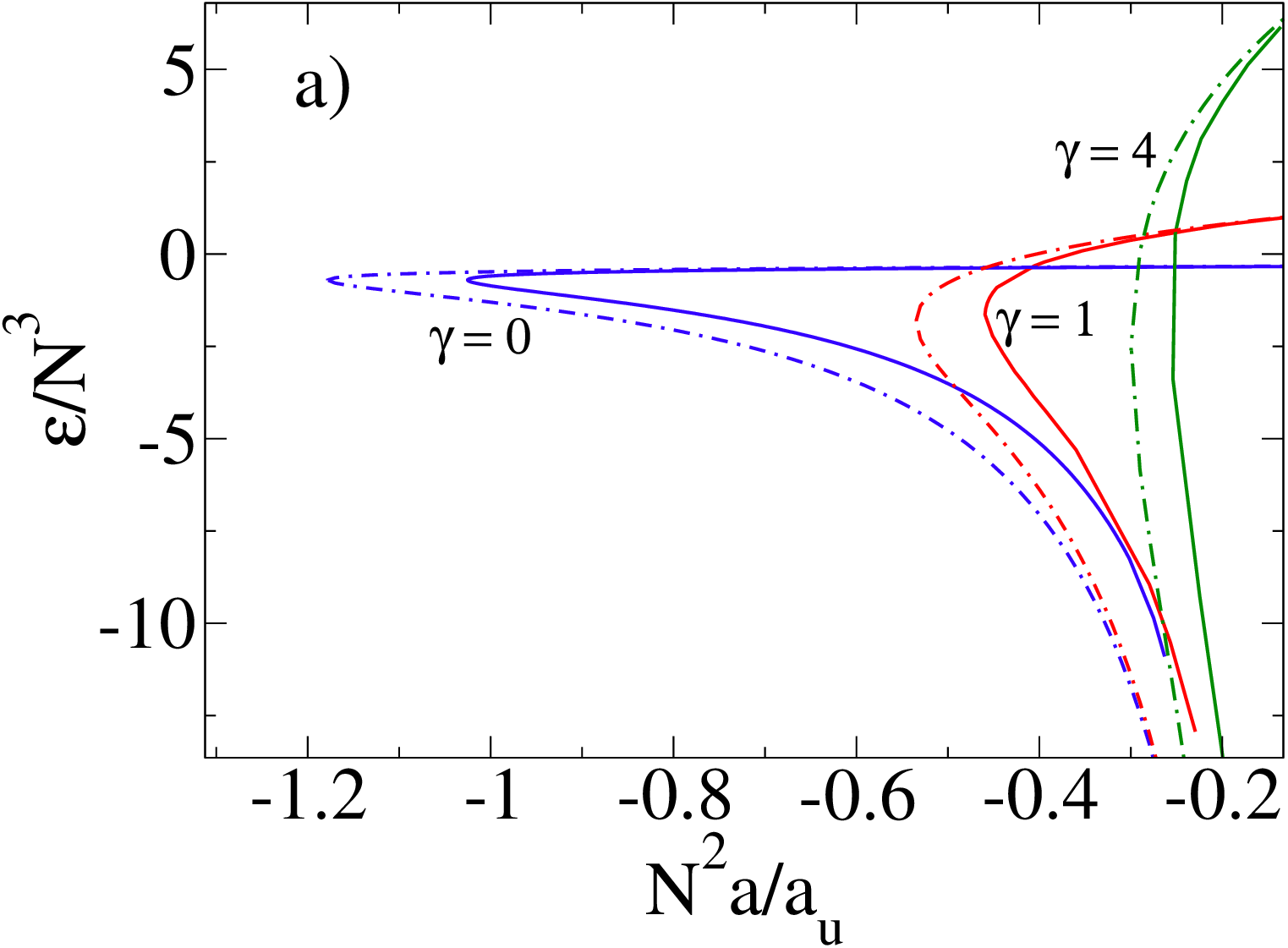}
\\[1ex]
{\includegraphics[width=0.39\textwidth]{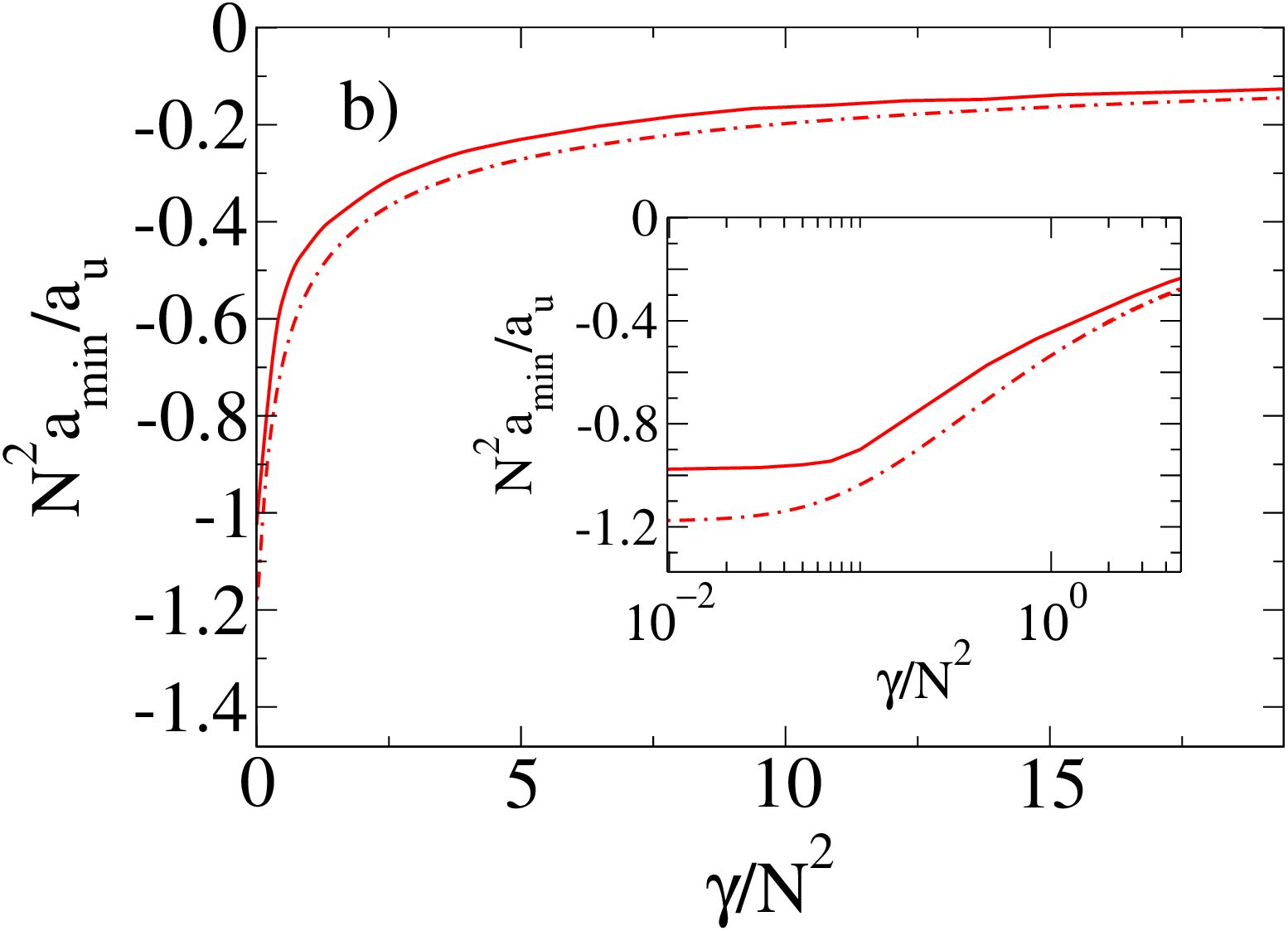} \hskip 0.0 cm}
\caption{\label{fig7} (Color online) a) Bifurcations of the chemical potential for  nonvanishing values of the  trapping potential. The case $\gamma = 0$ is shown for comparison. b) Dependence of the critical scattering length (the bifurcation point) on the frequency of the trapping potential. Numerically accurate results are given by solid lines, variational results by dashed lines.
To elucidate the behavior for small values of $\gamma/N^2$, this region is  shown in the  inset on a logarithmic scale.}
\end{figure}
\section{Bifurcating solutions}
A new result of our numerical calculations is that for negative scattering lengths there exists a second
radially symmetric nodeless solution of the extended Gross-Pitaevskii equation (\ref{HFat}). 
In Fig.~\ref{fig6} the chemical potential of the two solutions are plotted as functions of the
scaling parameter $N^2 a/a_u$ for $\gamma = 0$. It is evident that the critical value of 
$N^2 a/a_u = -1.0251$ corresponds to a bifurcation point of the eigenvalue spectrum of  equation (\ref{HFat}): below the critical point no solution exists, at the critical
point two solutions appear in a tangent bifurcation. The total energies of the condensates corresponding
to the two solutions are also shown in  Fig.~\ref{fig6}.  It can be seen that the energy increases
from the bifurcation point on for the second solution. This would mean that above the bifurcation point 
there exists a {\em collective excited state} of the condensate in which all atoms occupy one and the same nodeless orbital, just like in the true ground state.

The second solution is in fact present also in the variational
calculation. It there appears as a second stationary (maximum) point of the mean-field energy given as  a function of 
the width of the Gaussian type orbital. The variational results for the chemical potential and the total
energy of the second solution are also included in Fig.~\ref{fig6}. It can be seen that  the numerically accurate
calculation is necessary for the quantitative description of the bifurcation.

The second solution even persists in 
external trapping potentials, for any value of $\gamma$. The bifurcation diagram for two finite values of $\gamma$ is shown 
in Fig.~\ref{fig7}. It can be seen that with growing $\gamma$ the bifurcation point is shifted to smaller absolute values of $N^2a/a_u$. The increase of the total energy of the second solution which is evident from the figures is a consequence of the fact that the self-consistent potentials become more and more binding and the wave functions more and more localized which leads to a dramatic increase in the kinetic energy. 

What is the physical meaning of the second solution? 

We note, on the one hand, that it corresponds to a maximum of the mean-field energy functional. Schr\"odinger's equation, however, and in our case equation (\ref{HFat}), follows as the Euler-Lagrange equation of a  variational principle which only demands the energy functional to be an extremum. Thus the fact that the second solution corresponds to a maximum of the energy functional does not preclude it from corresponding to a real physical quantum state. On the other hand, the two solutions are nodeless, and hence nonorthogonal.  Obviously this is a
consequence of the nonlinearity of the extended Gross-Pitaevskii equation (\ref{HFat}): each solution creates its own self-consistent
potential and thus sees a different Hamiltonian. This would seem surprising since the original many-body Hamiltonian
is Hermitian and linear in the wave function, and therefore should possess only orthogonal eigenstates. The nonlinearity of (\ref{HFat}) is a result of the Hartree approximation made for the states.

In  studies of the decay rates in attractive trapped Bose-Einstein condensates,  with contact interaction only, Huepe et al. \cite{huepe99, huepe03} have 
seen similar behavior, i.~e., a second solution is born in a tangent bifurcation together with the ground state. These states also are nonorthogonal.
Analyzing the stability of the states Huepe et al. have shown that the first excited state out of the two solutions is unstable with respect to macroscopic quantum tunneling. 

This is a strong indication that the second solution found in this paper in Bose condensates with gravity-like interaction also corresponds to an unstable collectively excited state. A way to establish this is to linearize
the time-dependent Gross-Pitaevskii equation corresponding to (\ref{HFat})
around the stationary states
and to carry out  a stability analysis, as was done for the case of 
a pure attractive contact interaction by Huepe et al. \cite{huepe99, huepe03}.
Alternatively, by choosing a Gaussian ansatz with time-dependent widths \cite{garcia97},
equations of motion for the widths  can be obtained from the time-dependent Gross-Pitaevskii equation and analyzed with standard stability methods 
of nonlinear dynamics. Investigations along these lines are under way.
 
We finally note that there is an analogy with bifurcations seen in investigations
of attractive one-dimensional Bose-Einstein condensates on a ring (cf., e.\ g.,  \cite{carr00,kanamoto03,alon04}). There, at a critical
value of the ratio of the mean-field interaction energy to the kinetic energy,
symmetry-breaking, soliton-like solutions appear, in addition to the symmetry-preserving solution of the Gross-Pitaevskii equation, which are lower
in energy. By contrast, in the example discussed in this paper, both bifurcating
solutions possess the same symmetry.

\section{Conclusions}
We have re-analyzed Bose condensates with attractive $1/r$ interaction by introducing appropriate atomic units which are in particular adapted to the case of self-binding. We have thus been able to derive new scaling properties of such condensates. We have calculated numerically accurate results for wave functions and observables of self-binding condensates and compared them with previous variational results. It turned out that in particular at negative scattering lengths the variational results become poor and have to be replaced with our accurate numerical results. As a novel finding we have demonstrated that the critical point where collapse of the condensate occurs at negative scattering lengths is in reality a bifurcation point of the energy functional where both the ground state and an excited state merge and disappear. We have argued that this second solution indicates the
existence of an unstable collectively excited state at negative scattering lengths in degenerate Bose condensates with long-range attractive $1/r$ interaction.

Critical points, below which collapse of the condensate sets in,  not only
exist in attractive condensates at negative scattering lengths \cite{garcia97,huepe99, huepe03,carr00,kanamoto03,alon04} and in the
monopolar  gases with gravity-like interaction discussed in this paper but also exist in dipolar gases, in certain parameter ranges of the particle number, the scattering length and the trap frequencies \cite{santos00,goral00,dutta07}. Our investigations
suggest that these also correspond to bifurcation points. Studies of 
the bifurcation scenarios in dipolar gases are therefore strongly encouraged.
\begin{acknowledgments}
We thank Axel Pelster for useful dicussions.
\end{acknowledgments}


\end{document}